\documentclass[prb,reprint,amsmath,amssymb,showpacs,superscriptaddress,longbibliography,floatfix]{revtex4-1}
\usepackage{graphicx,epsfig}
\usepackage{amsmath}
\usepackage{xcolor}
\usepackage[breaklinks=true,colorlinks,citecolor=blue,linkcolor=blue,urlcolor=blue]{hyperref}
\usepackage{subfigure}
\usepackage{color}
\def\be{\begin{equation}}
\def\ee{\end{equation}}

\def\bea{\begin{eqnarray}}
\def\eea{\end{eqnarray}}

\begin{document}

\title{Broadband excitation spectrum of bulk crystals and thin layers of PtTe$_2$}
\author{Barun Ghosh}
\affiliation{Department of Physics, Indian Institute of Technology Kanpur, Kanpur - 208016, India}
\author{Francesca Alessandro}
\affiliation{Department of Physics, University of
Calabria, via ponte Bucci, cubo 31/C 87036, Rende (CS) Italy}
\affiliation{INFN, Sezione LNF, Gruppo Collegato di Cosenza, Cubo 31C, I-87036 Rende (CS), Italy}
\author{Marilena Zappia}
\affiliation{Department of Physics, University of
Calabria, via ponte Bucci, cubo 31/C 87036, Rende (CS) Italy}
\author{Rosaria Brescia}
\affiliation{Electron Microscopy Facility, Istituto Italiano di Tecnologia, Via Morego 30, 16163 Genova, Italy}
\author{Chia-Nung Kuo}
\affiliation{Department of Physics, National Cheng Kung University, 1 Ta-Hsueh Road 70101 Tainan, Taiwan}
\author{Chin Shan Lue}
\affiliation{Department of Physics, National Cheng Kung University, 1 Ta-Hsueh Road 70101 Tainan, Taiwan}
\author{Gennaro Chiarello}
\affiliation{Department of
Physics, University of Calabria, via ponte Bucci, cubo 31/C 87036,
Rende (CS) Italy}
\author{Antonio Politano}
\affiliation{Istituto Italiano di Tecnologia-Graphene Labs via Morego, 30 16163 Genova, Italy}
\affiliation{Dipartimento di Scienze Fisiche e Chimiche (DSFC), Universita dell'Aquila, Via Vetoio 10, I-67100 L'Aquila, Italy}
\author{Lorenzo S. Caputi}
\affiliation{Department of Physics, University of
Calabria, via ponte Bucci, cubo 31/C 87036, Rende (CS) Italy}
\author{Amit Agarwal}
\email{amitag@iitk.ac.in}
\affiliation{Department of Physics, Indian Institute of Technology Kanpur, Kanpur - 208016, India}
\author{Anna Cupolillo}
\email{anna.cupolillo@fis.unical.it}
\affiliation{Department of Physics, University of
Calabria, via ponte Bucci, cubo 31/C 87036, Rende (CS) Italy}
\date{\today}

\begin{abstract}
We explore the broadband excitation spectrum of bulk PtTe$_2$ using
electron energy loss spectroscopy and density functional theory.
In addition to infrared modes related to intraband 3D Dirac
plasmon and interband transitions between the 3D Dirac bands, we
observe modes at 3.9, 7.5 and 19.0 eV in the ultraviolet region. The
comparison of the excitation spectrum with the calculated
orbital-resolved density of states allows us to ascribe spectral
features to transitions between specific electronic states.
Additionally, we study the thickness dependence of the high-energy
plasmon in the PtTe$_2$ thin films. We show that, unlike graphene, the high-energy plasmon in PtTe$_2$ thin film gets red-shifted by $\sim$2.5 eV with increasing thickness.
\end{abstract}

\maketitle
\section{Introduction}

Recently, the PtX$_2$ (X=Se, Te, S) class of transition-metal
dichalcogenides (TMDCs) have attracted a huge interest of the
scientific community. This class of TMDCs combines promising  
application capabilities along with the fundamental physics interest arising from the
existence of topological type-II Dirac fermions. \citep{Castellanos-Gomez2016,doi:10.1002/adfm.201701011,Soluyanov2015,Yan2017,Bahramy2017,PhysRevLett.119.016401,PhysRevB.96.125102,PhysRevB.94.121117,Yan2017,PhysRevLett.119.026404} As opposed
to type-I Dirac materials, which have a closed Fermi surface with
either an electron or a hole pocket, type-II Dirac materials have
an unbounded Fermi surface with both electron and hole pockets. \citep{Soluyanov2015}
The presence of bulk topological Dirac node forces the existence
of massless surface states. This provides massless charge carriers
with ultrahigh mobility confined at the surface plane. \citep{Bahramy2017}

Thin layers of PtX$_2$ class of materials are equally interesting because of their air stability, high mobility, and superior gas sensing properties, among others. \cite{doi:10.1002/adfm.201701011,Lin2017,zhao2017high,chia2016layered,doi:10.1021/acs.nanolett.5b00964,ADMA201602889} The quantum confinement effect dramatically changes the properties of the thin layers of PtTe$_2$ which undergo a metal-semiconductor transition with decreasing thickness. \citep{doi:10.1021/acs.nanolett.8b00583,Ciarrocchi2018} Apart from this, the monolayer exhibits a unique property: because of local dipole-induced Rashba effect, opposite spins with the same energy gets spatially separated on the opposite sides of the monolayer. \cite{Ya2017} This is known as spin layer locking and it can have potential applications in the electrically tunable spintronic devices.

Preceding discussions made it clear that PtX$_2$ class of materials has emerged as a promising material for the future electronics. While the band structure of PtTe$_2$  has been explored comprehensively \citep{Bahramy2017}, along with the Dirac plasmons (collective density excitations) in the infrared range of the electromagnetic spectrum \cite{PtTe2PRL}, the high-energy excitations in PtTe$_2$  still remain unexplored. The comprehension of the excitation spectrum of collective modes in the visible-ultraviolet is crucial to devise broadband photodetectors \citep{Koppens2014,Yu2018}, ultraviolet-imaging applications \citep{7089216} and broadband plasmonic devices. \citep{Ou2014,doi:10.1021/nn503035b} Monolayer PtTe$_2$ has the smallest energy band gap in the PtX$_2$ class of materials\citep{0268-1242-31-9-095011}, offering it an advantage over the other members for applications in nano-electronics. 

Motivated by this, in this paper we probe the broadband excitation spectrum of
bulk crystals and thin layers of PtTe$_2$, using electron energy loss spectroscopy (EELS)
\cite{ROCCA19951} complemented by detailed {\it ab-initio}
calculations. EELS probes the broadband dielectric response of the
system to a negatively charged probe, allowing for spectral
contributions from both plasmonic modes and non-vertical
transitions from valence-band to conduction-band electronic
states. Therefore, EELS investigations supplement  the studies of
the absorption and emission processes of TMDCs in the
long-wavelength limit \citep{C4CS00265B} involving only vertical
transitions from occupied to unoccupied states.
\begin{figure*}[t]
\includegraphics[width=.75\linewidth]{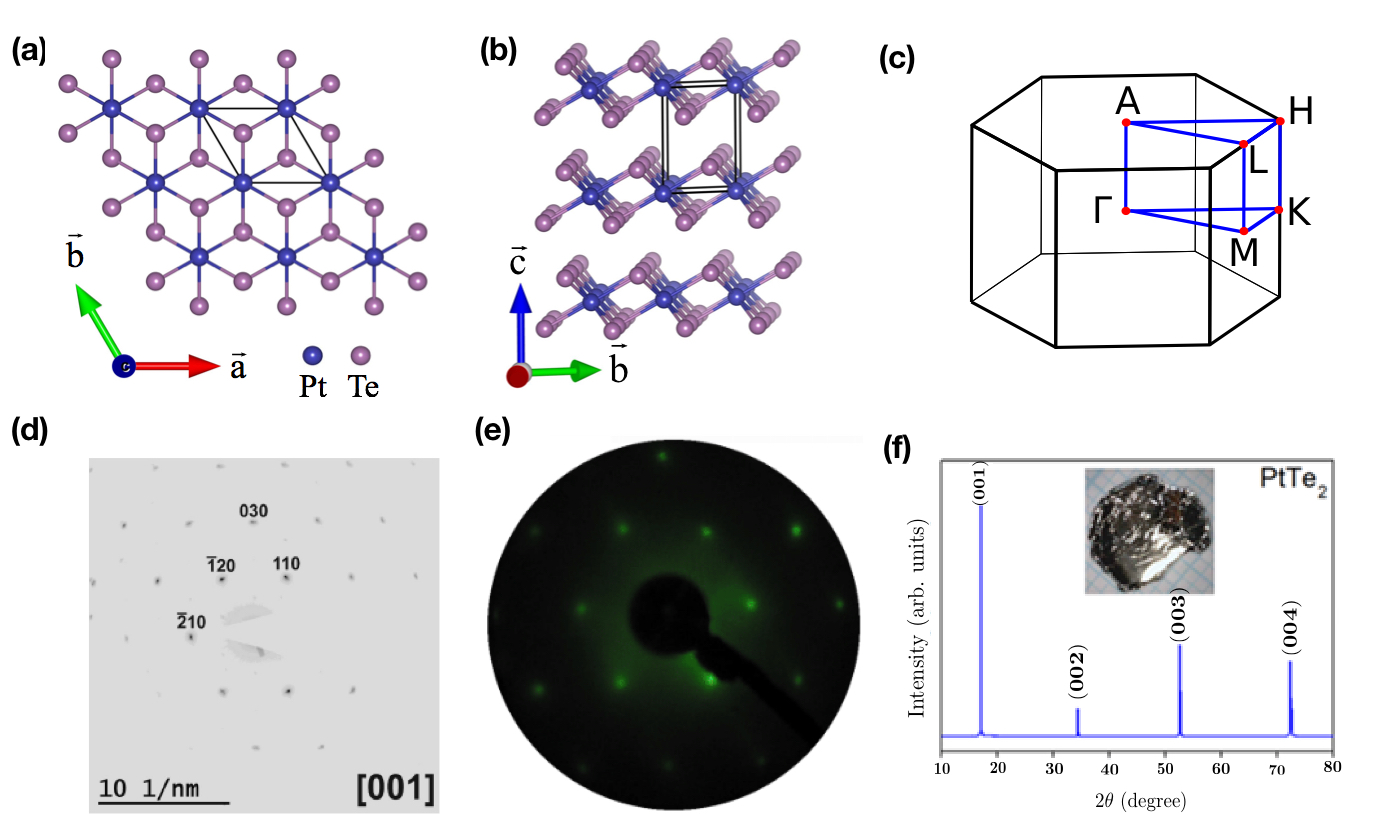}
\caption{Panels (a) and (b) show the top and side view of the
atomic crystal structure of PtTe$_2$. The corresponding bulk
Brillouin zone of the hexagonal PtTe$_2$ crystal, along with
various high symmetry points is shown in (c). (d)
Selected area electron diffraction (SAED) patterns acquired on PtTe$_2$ flakes match with [0001]-oriented single-crystal trigonal PtTe$_2$ (ICSD $\#41373$, moncheite).
(e) LEED pattern of bulk PtTe$_2$ single crystal oriented along the (0001) direction. (f) XRD pattern of (0001)-oriented planes of PtTe$_2$. Only (00n) peaks are observed.
\label{fig1}}
\end{figure*}
%
The nature of various experimentally probed EELS excitations
is identified by means of 
calculations of the band-structure and the loss function based on density-functional theory (DFT).
In addition to the bulk excitation spectrum, we also explore the
changes in the broadband excitation spectrum of PtTe$_2$ on
reducing its thickness to few layers. Contrarily to the case of
graphene and phosphorene, a blue-shift of the high-energy plasmon
frequency is observed for thin layers compared to the
bulk.\citep{PhysRevB.77.233406,PhysRevB.90.235434,PhysRevB.96.035422}

The manuscript is organized as follows: In Sec.~\ref{sec2} we
briefly describe the crystal structure, experimental details related to the growth
and characterization of PtTe$_2$ along with that of the reflection EELS. This is followed by a discussion of the theoretical {\it ab-initio} calculation of the EELS spectrum and  other computational details in Sec.~\ref{sec3}.  The broadband EELS spectrum of bulk PtTe$_2$ is discussed in Sec.~\ref{sec4}, followed by the discussion of the
excitation spectrum of thin layers in Sec.~\ref{sec5}.
Finally, we summarize our findings in Sec.~\ref{Conc}.

\section{Growth, Crystal structure, Characterization and details of EELS}
\label{sec2}
\subsection{Growth}
Single crystals of PtTe$_2$ were prepared by the self-flux method. High-purity Pt (99.99 \%) foil and Te ingot (99.9999 \%) were mixed in the ratio of 1:17 and wrapped in a quartz tube under vacuum. The quartz tube was heated to 1000 $^\circ$C, dwelled there for 8 hours, and slowly cooled at a rate of 3-5 $^\circ$C/h to 500 $^\circ$C. Successively, the excess Te flux was separated by centrifugation. The resulting crystals have typical dimensions of $8  \times 8 \times 1$ mm \citep{zhao2017high} with the c-axis perpendicular to the plates and can be easily cleaved. The structure of the grown crystals was examined by X-ray diffraction (Bruker D2 PHASER) using Cu K$_\alpha$ radiation and Laue diffraction at room temperature.

{Thin flakes of PtTe$_2$ were obtained by liquid-phase exfoliation of bulk PtTe$_2$ in N-methylpyrrolidone (NMP).} 

\subsection{Crystal Structure}
Bulk PtTe$_2$ belongs to the large family of 1T-metal dichalcogenides with CdI$_2$ type crystal structure [space group P$\bar{3}$m1(164)]. The bulk structure can be viewed as a collection of isolated monolayers stacked in the out of plane direction -- see Fig.~\ref{fig1}(a) and ~\ref{fig1}(b). In each of these monolayers there are three sublayers, Te-Pt-Te, where the central Pt atom is strongly bonded with six neighbouring Te atoms forming a hexagonal honeycomb structure. Both selected area electron diffraction (SAED) and low-energy electron diffraction (LEED) patterns shown in Fig.~\ref{fig1}(d) and \ref{fig1}(e)  match with [0001]-oriented flakes and bulk PtTe$_2$, respectively. 
Correspondingly, the XRD pattern only exhibits (00n) peaks as shown in Fig.~\ref{fig1}(f). 

\subsection{Characterization}
To demonstrate cleanliness, the surface has been
characterized by means of vibrational spectroscopy and X-ray
photoelectron spectroscopy, without finding any contamination (see appendix B, and C).
Once prepared in ultra-high vacuum, the surface remains
uncontaminated for a timescale of several weeks, thus ensuring
sample stability. The low-energy electron diffraction (LEED)
pattern shows sharp spots against a low background -- see Fig.~\ref{fig1}(e).

\subsection{Electron Energy Loss Spectroscopy}
The reflection EELS experiments were performed at room temperature
by means of an EELS apparatus with two 50 mm hemispherical
deflectors for both monochromator and analyzers, mounted in an
ultra-high vacuum chamber at the University of Calabria, Italy. The
primary electron beam impinges on the sample with an incident
angle $\theta_i$ of 43$^\circ$ with respect to the surface normal,
along the $\Gamma-K$ direction of the surface Brillouin zone. The
primary electron beam energy is $E_p = 100$ eV.

SAED and STEM-EELS analyses were carried out at room temperature on a FEI Tecnai G2 F20 TWIN TEM, equipped with a Gatan Enfinium SE spectrometer at Istituto Italiano di Tecnologia, Genoa (Italy). For these experiments a primary electron beam energy $E_p = 200$ keV and a collection angle of 13 mrad were used. Samples for TEM analyses were prepared by drop casting of the flakes dispersion onto a holey carbon-coated Cu grid. The EELS spectra were collected from flakes regions suspended on holes in the carbon support film.


\begin{figure*}[t]
\includegraphics[width=0.95\linewidth]{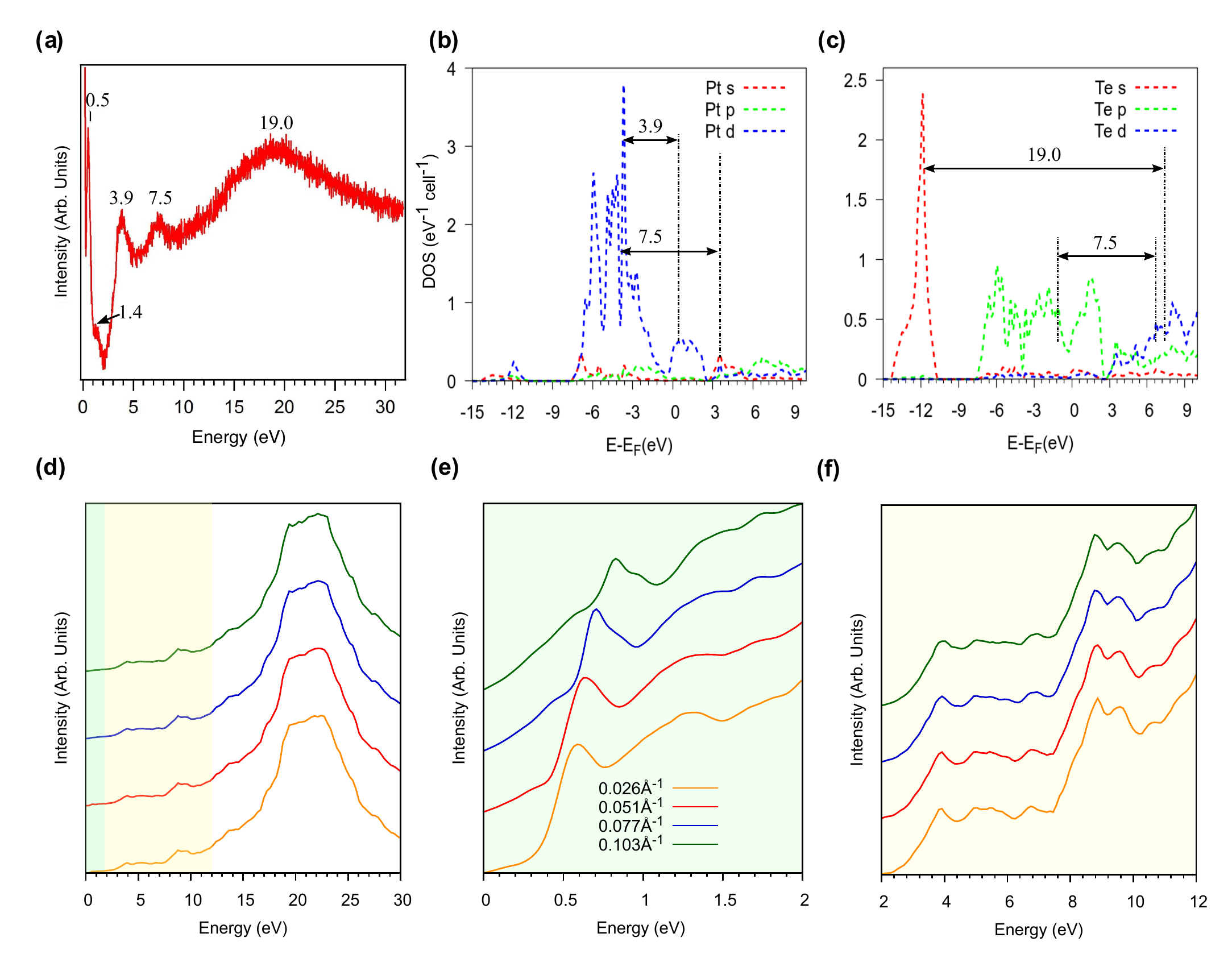}
\caption{(a) Broadband EELS spectrum for bulk PtTe$_2$ measured in
reflection mode with a primary electron beam energy of 100 eV. The
EELS spectrum shows several distinct peaks at energies 0.5, 1.4,
3.9, 7.5 and 19.0 eV. The peak at 0.5 eV is dispersive and it is
associated with the intra-band  density excitations
\cite{PtTe2PRL} in PtTe$_2$.  The other higher energy peaks
correspond to inter-band transitions, which can be identified from
the orbital-resolved DOS plot, shown in the panels
(b) and (c). The corresponding orbital-resolved band-structure
plots are shown in Fig.~\ref{fig3}. The dominant transitions,
corresponding to the observed EELS peaks, are marked by arrows.
The experimental broadband EELS spectrum is also reasonably
captured by the loss function obtained from {\it{ab-initio}}
calculations, as shown in (d)-(f) for different momentum values, reported in the legend of panel (e).
The 19.0 eV peak in (d) has the highest intensity in the {\it ab-initio}
calculations. The other dominant intra-band peaks at 3.9 and 7.5
eV are resolved in panel (f).} \label{fig2}
\end{figure*}

\section{Theory}
\label{sec3}
\subsection{Energy Loss function}
The theoretical calculation of the electron energy loss function
starts with the non-interacting density-density response function
$( \chi^0_{\bf{GG'}} )$ for a periodic lattice. It is obtained
using the  Adler-Wiser formula given
by\cite{PhysRev.126.413,PhysRev.129.62},
\begin{eqnarray} \label{eq:chi0}
& & \chi^0_{\mathbf{G} \mathbf{G}^{\prime}}(\mathbf{q}, \omega) = \frac{1}{\Omega}
\sum_{\mathbf{k}}^{\mathrm{BZ}} \sum_{n, n^{\prime}}
\frac{f_{n\mathbf{k}}-f_{n^{\prime} \mathbf{k} + \mathbf{q} }}{\omega + \epsilon_{n\mathbf{k}} - \epsilon_{n^{\prime} \mathbf{k} + \mathbf{q} } + i\eta}  \times \\
& & \langle \psi_{n \mathbf{k}} | e^{-i(\mathbf{q} + \mathbf{G}) \cdot \mathbf{r}} | \psi_{n^{\prime} \mathbf{k} + \mathbf{q} } \rangle_{\Omega_{\mathrm{cell}}}
\langle \psi_{n\mathbf{k}} | e^{i(\mathbf{q} + \mathbf{G}^{\prime}) \cdot \mathbf{r}^{\prime}} | \psi_{n^{\prime} \mathbf{k} + \mathbf{q} } \rangle_{\Omega_{\mathrm{cell}}}~. \nonumber
\end{eqnarray}
The wave function $\psi_{n\bf{k}}$, eigenvalue $\epsilon_{n\bf{k}}$, and the corresponding Fermi-Dirac occupation function $f_{n\bf{k}}$ at wave vector $\bf{k}$ for the band with index $n$ are obtained from the ground-state calculations performed using DFT.

\begin{figure*}
\includegraphics[width=0.9\linewidth]{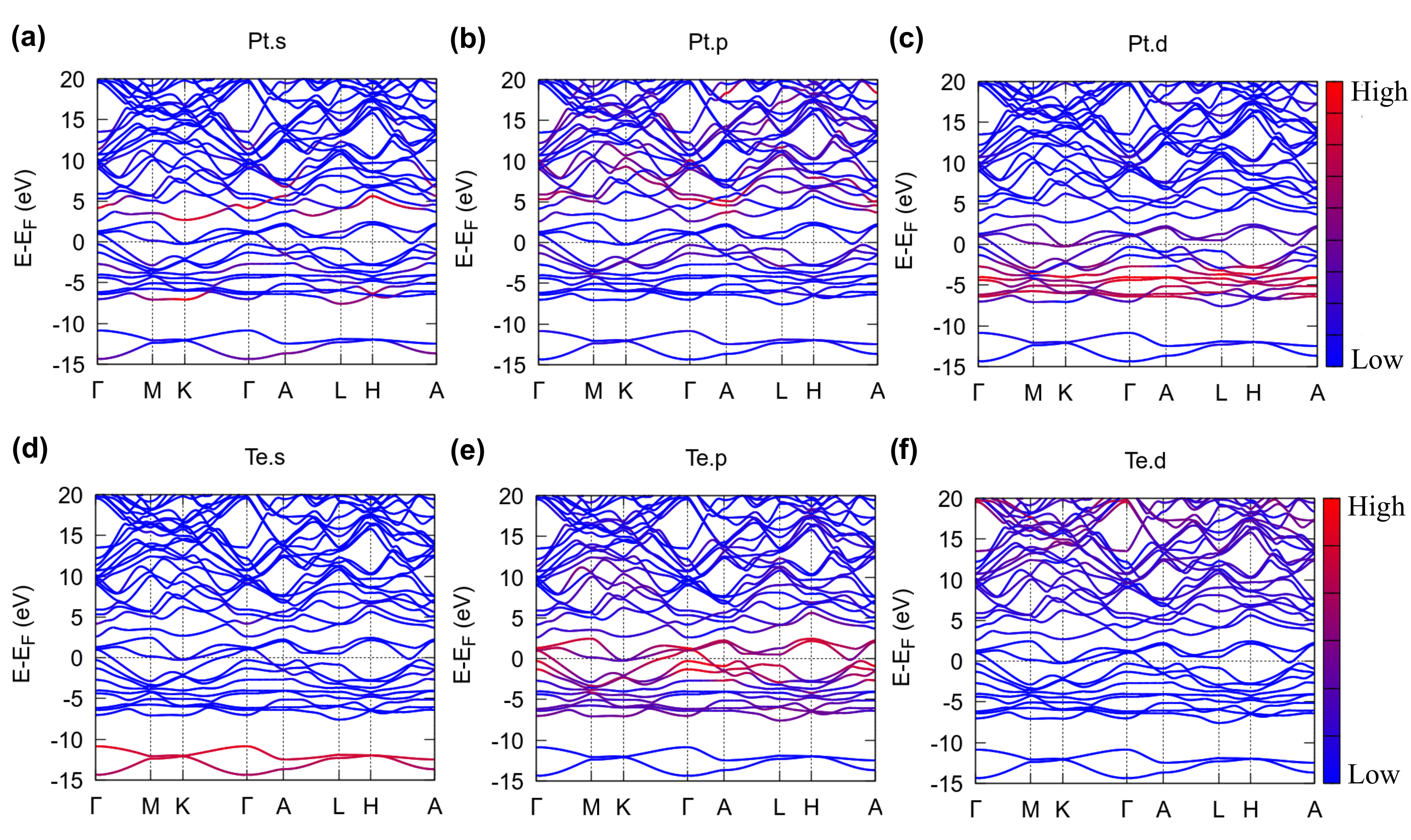}
\caption{Orbital-projected band structure of PtTe$_2$. The lowest
energy states are dominated by Te$_{5s}$ orbital. Near the Fermi
energy the states have major contributions from Pt$_{5d}$  and
Te$_{5p}$ orbitals. The higher energy states are mostly Pt$_{6p}$
and Te$_{5d}$ states.} \label{fig3}
\end{figure*}

The interacting density-density response function can be obtained
within the framework of time-dependent DFT 
(TDDFT) by solving a Dyson-like equation. Using a plane-wave basis
for a periodic system, for a given $(\bf{q},\omega)$, it can be
expressed as
\begin{equation}
\chi_{\bf{GG'}} = \chi_{\bf GG'}^0
+ \sum_{\bf{G_1,G_2}}\chi_{\bf{GG_1}}^0({\bf q},\omega) K_{\bf{G_1G_2}}(q)\chi_{\bf{G_2G'}}(\bf{q},\omega), \nonumber
\end{equation}
where $\bf{G}$ and $\bf{q}$ are the reciprocal lattice vector and
the Bloch wave vector, respectively, and $K_{\bf{G_1G_2}}$ is the
interaction kernel, including both the Coulomb or Hartree term, as
well as the exchange correlations. 

Using the calculated 
$\chi_{\bf{G}\bf{G'}}$ matrix, the dielectric matrix, $\epsilon_{\bf{G}\bf{G'}}({\bf q},\omega)$, can obtained as,
\begin{equation}
\epsilon^{-1}_{\mathbf G \mathbf G^{\prime}}(\mathbf q, \omega)
= \delta_{\mathbf G \mathbf G^{\prime}}+\frac{4\pi}{|\mathbf q+\mathbf G|^2}
\chi_{\mathbf G\mathbf G^{\prime}}(\mathbf q, \omega)~.
\end{equation}
The macroscopic dielectric constant can be obtained from the dielectric matrix via the equation, 
\begin{equation}
\epsilon_M(\mathbf q, \omega) = \frac{1}{\epsilon^{-1}_{00}(\mathbf q, \omega)}~.
\end{equation}
The dynamical loss function, which is directly related to the experimental excitation spectrum probed by EELS, is now obtained from the macroscopic dielectric function, 
\begin{equation}
{\cal E}_{\rm Loss}({\bf q},\omega)=-\Im\left[\frac{1}{\epsilon_M(\mathbf q,\omega)}\right]~.
\end{equation}

Plasmons (collective density excitations) are characterized by the
zeroes of the real part of the macroscopic dielectric function
(the denominator of the density-density response function within
RPA). Thus, they appear as peaks in the EELS spectrum. The
calculations of the loss function are performed using the
generalized RPA, which uses the local-field factors to add the
impact of the exchange and correlation effects, in addition to the
Hartree field \cite{GV}.

\subsection{Computational details}
Electronic band structure calculations were performed using a
plane-wave basis set within the framework of DFT, as implemented in the VASP
package.\citep{PhysRev.140.A1133,PhysRevB.54.11169} We use the PAW
pseudopotentials with 500 eV kinetic energy cut-off for the
plane-wave basis set.\citep{PhysRevB.59.1758}  The
exchange-correlation part of the potential has been treated within
the framework of generalized approximation scheme developed by
Perdew-Burke-Ernzerhof.\citep{PhysRevLett.77.3865} Starting from
the experimental structure, we relax all the atomic positions
until the forces on each atom are less than 0.001 eV/\AA.
Spin-orbit interaction has been considered as a perturbation and
treated in a self-consistent manner.

In order to calculate the response functions we use the GPAW code,
which employs a real-space representation of the PAW potentials.
\cite{PhysRevB.71.035109,0953-8984-22-25-253202,ISI:000175131400009}
A kinetic energy cutoff of 500 eV has been used for the plane-wave
basis set. We use a 122$\times$122$\times$20 $k$-grid to calculate
the momentum dependence of the loss function. To incorporate the
local-field effects, we use a plane-wave cutoff of 60~eV which
corresponds to 85 plane waves. A broadening
parameter $\eta=0.05$~eV is assumed in all calculations of the
response function.

\section{Broadband spectrum of bulk $\rm\bf{{PtTe_2}}$}
\label{sec4}

The experimental broadband EELS spectra of PtTe$_2$ bulk sample,
measured with an electron beam energy of 100 eV, is shown in
Fig.~\ref{fig2}(a). The broadband EELS spectrum shows distinctly
resolved peaks at energies $\sim0.5$, $\sim 1.4$, $\sim3.9$, $\sim 7.5$ and $\sim 19.0$ eV, among
others. The lowest energy peak at 0.5 eV is the intra-band gapped
3D Dirac plasmon excitation in bulk PtTe$_2$, which disperses with
the momentum [see Fig.~\ref{fig2}(e)]. It has been discussed in
detail in Ref.~[\onlinecite{PtTe2PRL}]. Here, we will focus on the
remaining inter-band excitations, which are relatively less
dispersive compared to the intra-band 3D Dirac plasmon peak. We
find that their peak location and relative intensity is nearly
independent on scattering geometry and, consequently on
the momentum in a momentum range of of 0.0-0.2
\AA$^{-1}$.

In order to identify the inter-band transitions corresponding to
the observed peaks in the EELS spectrum, we show the
orbital-resolved density of states (DOS) in
Fig.~\ref{fig2}(b)-(c). The corresponding orbital-resolved band
structure plot is shown in Fig.~\ref{fig3}.   The energetically
lower valence band (VB) is situated approximately between -15 and -10 eV and it
is mostly dominated by Te$_{5s}$ states. The upper VB extends from
-7 eV up to the Fermi level and mostly comprises of Pt$_{5d}$ and
Te$_{5p}$ orbitals. The lower conduction-band states till $\sim$3
eV are also primarily formed by Pt$_{5d}$ and Te$_{5p}$ orbitals.
The conduction-band states at higher energy have contributions
mainly from the Pt$_{6p}$ and Te$_{5d}$ orbitals.
Comparison of the orbital-resolved DOS with the observed spectral
features in the broadband EELS spectrum allows us to identify the
prominent states involved in the transitions.
We find that the peaks at 3.9, 7.5 and 19.0 eV are predominantly
connected to Pt$_{5d}~\to$ Pt$_{5d}$; Te$_{5p}~\to$ Te$_{5d}$; and
Te$_{5s} \to$ Te$_{5d}$ transitions, respectively, as marked by
arrows in Fig.~\ref{fig2}(b)-(c). The theoretical loss function in
Fig.~\ref{fig2}(d)-(f) also captures the qualitative features of
the experimental excitation spectrum reasonably well. 

It is worth reminding that EELS peaks cannot precisely match the energies of single-particle transitions observed with optical spectroscopies, as maxima in the loss function are related to the maxima in $-{\rm Im}[1/\epsilon_M(\omega)]$. Conversely, optical transitions are identified with maxima in  ${\rm Im}[\epsilon_M(\omega)$] \citep{wooten1935optical}, that are actually shifted with respect to the maxima of $-{\rm Im}[1/\epsilon_M(\omega)]$ \citep{liebsch2013electronic}. Having discussed the EELS spectrum of the bulk PtTe$_2$ crystal, we now proceed to discuss the EELS spectrum of thin PtTe$_2$ layers in the next section.

\begin{figure*}[t]
\includegraphics[width=.9\linewidth]{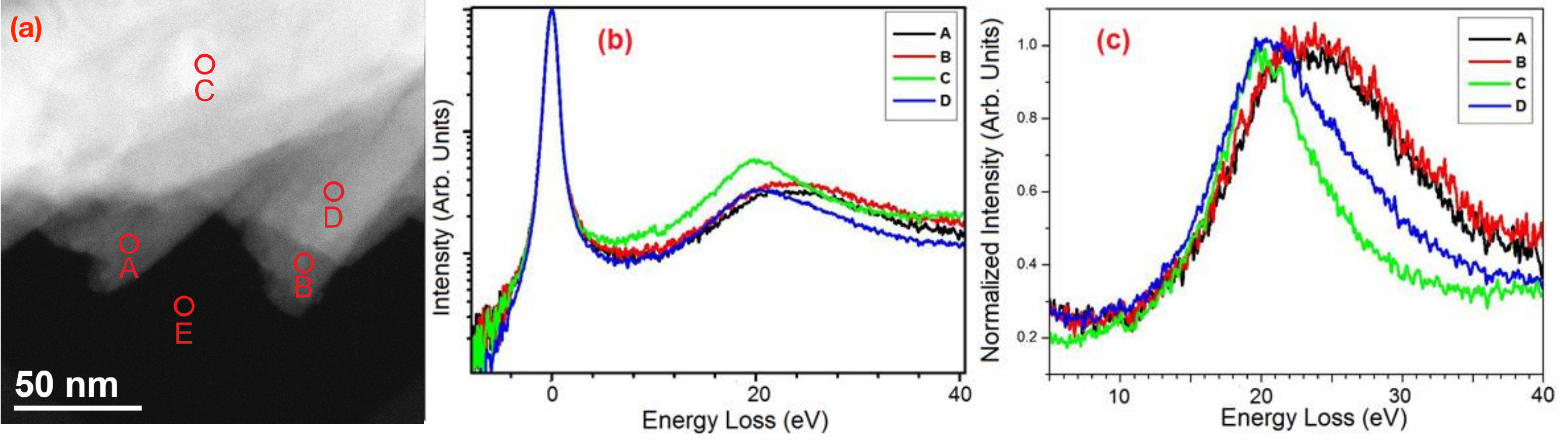}
\caption{(a) HAADF-STEM image of the fragmented PtTe$_2$ flake suspended on a hole in the amorphous carbon support film used for STEM-EELS analysis. 
The points A and B exhibit a lower thickness (lower brightness in HAADF-STEM mode) than point D, while the point C has the highest thickness in the flake. Only the zero-loss peak appears in the spectrum recorded at point E, i.e. 20 nm far from the PtTe$_2$ flake.
(b) The
broadband STEM-EELS spectra acquired for points A, B, C and D. 
In panel (b) the intensity is normalized to the zero-loss peak maximum, while in panel (c) it is normalized to that of the peak at 
$\sim$20 eV, in order to highlight the dissimilarities in the
line-shape of the peaks corresponding to different thickness.} \label{fig4}
\end{figure*}

\section{Broadband spectrum of thin $\rm\bf{{PtTe_2}}$ layers}
\label{sec5} In this section, we explore the thickness dependence of
the broadband EELS spectrum of PtTe$_2$ thin films. To this aim,
we have employed the STEM-EELS technique, which affords the spatial
resolution necessary to identify regions with different thickness within the flakes. As shown in
Fig.~\ref{fig4}(a), the liquid-phase exfoliation of PtTe$_2$ has
produced flakes with inhomogeneous thickness. Therefore, for STEM-EELS experiments different
thicknesses were probed on different areas of same fragments.

Previously, the thickness dependence of the high energy EELS peaks
has been done for the case of $\sigma +\pi$ plasmons of
graphene. In graphene, the plasmon energy shifts by about 10 eV
when going from monolayer graphene to multilayers. \citep{PhysRevB.80.113410,PhysRevB.77.233406,PhysRevB.90.235434}
Remarkably, a similar blue-shift has been detected
by both reflection EELS in graphene layers epitaxially grown on
silicon carbide \citep{PhysRevB.80.113410}, and by EELS-TEM in
free-standing graphene flakes exfoliated from graphite.
\citep{PhysRevB.77.233406,PhysRevB.90.235434} This is
irrespective of the dissimilarities in the plasmon energy in the
monolayer regime (10 eV in the latter
case\citep{PhysRevB.77.233406,PhysRevB.90.235434} and 14.6 eV in
the former one\citep{PhysRevB.80.113410}).  This blue-shift of the
plasmon energy in graphene with increasing thickness had been
ascribed to the effect of interlayer Coulomb coupling
\cite{PhysRevB.90.235434}. In contrast to graphene, the high energy plasmon peak in 
phosphorene does not show any thickness dependence \cite{doi:10.1116/1.4926753}. 

Our STEM-EELS investigation suggests that in PtTe$_2$ the plasmon
band changes in both line-shape and energy position as a function
of thickness. Specifically, a symmetrical line-shape, centered at
about 23 eV (Fig.~\ref{fig4}c), is recorded for thinner flakes,
whereas an asymmetrical line-shape, with a corresponding centroid
around 20.5 eV, are revealed for thicker flakes. Thus, in contrast
to graphene\cite{PhysRevB.77.233406,PhysRevB.90.235434}, and phosphorene\cite{doi:10.1116/1.4926753} the high energy plasmon peak in PtTe$_2$ is
red-shifted with increasing thickness. However, the
layer-dependent plasmon-energy shift in PtTe$_2$ is much lower
($\sim$2.5 eV) than that of graphene ($\sim$10 eV). 
The change in the line shape, with an asymmetric shape characterizing the thinner regions, is ascribable to the multiple contributions (i.e., due to regions with different thickness) to the spectra acquired in the thinner regions of the flakes. The presence of contributions from regions few nm away from the position of the electron beam is due to the delocalization of inelastic scattering, prominent in the low energy-loss range \cite{ahn2004transmission,egerton1996electron}. 
%

\section{Conclusions}
\label{Conc}

We have probed the broadband excitation spectrum of bulk crystals
and thin layers of PtTe$_2$, using EELS in reflection
mode for bulk and STEM-EELS for thin layers. In the case of bulk
PtTe$_2$ we find different modes in the ultraviolet regime at 3.9,
7.5 and 19.0 eV, in addition to the excitations associated to the
3D Dirac cones observed in the infrared range at 0.5 and 1.4 eV.
These observations are well explained by the DFT-based
orbital-resolved band-structure and DOS calculations.
Specifically, we find that in bulk PtTe$_2$ the observed peaks at
3.9, 7.5 and 19.0 eV are predominantly connected to Pt$_{5d}~\to$
Pt$_{5d}$; Te$_{5p}~\to$ Te$_{5d}$; and Te$_{5s} \to$ Te$_{5d}$
transitions, respectively. In thin layers, with decreasing thickness, the
high-energy plasmon peak gets shifted from 20.5 to 23.0 eV. This
red-shift with increasing thickness is in contrast to the
blue-shift observed in case of graphene.  Moreover, with
increasing number of layers an increase in the anisotropy of the
line-shape of the high-energy plasmon peak is observed. 
This peculiarity can be exploited for characterizing the thickness of PtTe$_2$ 
thin films. Similar physics is expected to play out in other members of the family, including 
PdTe$_2$ and PtSe$_2$, among others.

\appendix

\begin{figure}[b]
\includegraphics[width=1.0\linewidth]{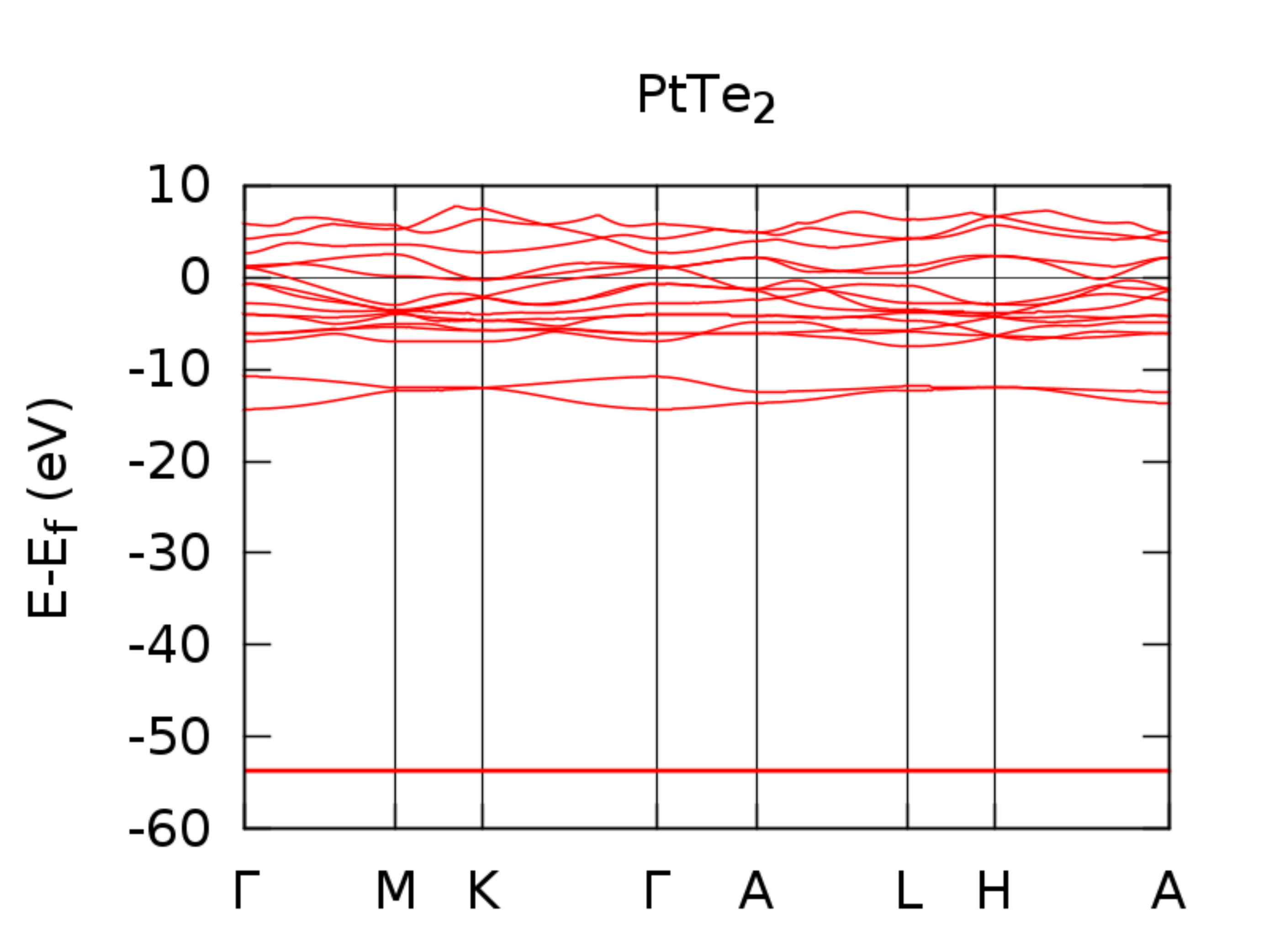}
\caption{Band structure of PtTe$_2$ over a wide energy region. Evidently, there are no energy states in between -13 eV to -50 eV.}
\label{supp_1}
\end{figure}

\section{Valence-band states in PtTe$_2$}
As an additional check we explicitly check that there are no
valence-band states below the ~-14 eV in PtTe$_2$, by calculating
the band-structure over a wide energy range. The calculated
band-structure is shown in Fig.~\ref{supp_1}: it is evident that
there are no valence-band states in PtTe$_2$ from approximately ~-14 to - 50 eV.


\begin{figure}[t]
\includegraphics[width=.8\linewidth]{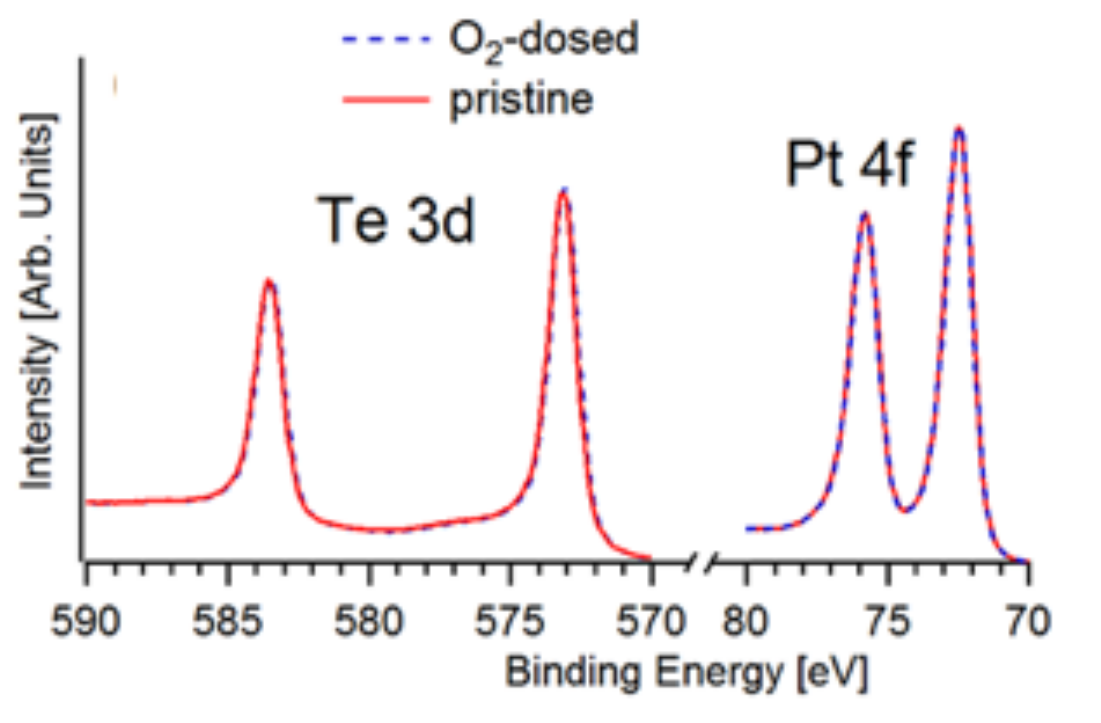}
\caption{High-resolution XPS spectra of Pt-4f and Te-3d core levels taken for the pristine PtTe$_2$ sample (red curve) and the same surface exposed to $10^6$ L of O$_2$ (dashed blue curve). For all spectra, the photon energy is 745 eV and the energy resolution is 0.1 eV. }
\label{S3}
\end{figure}

\section{Surface chemical reactivity of PtTe$_2$ single crystals}
To evaluate the surface chemical reactivity of PtTe$_2$, we have carried out high-resolution X-ray photoelectron spectroscopy (XPS) experiments (Fig.~\ref{S3}). We addressed firstly the evolution of Pt 4f and Te 3d core levels upon several treatments (O$_2$ dosage and air exposure), with respect to spectra measured for pristine PtTe$_2$. 
From the analysis of Te 3d core-level spectra (Fig.~\ref{S3}), we deduce that the as-cleaved undefected PtTe$_2$ surface is inert to oxygen exposure. In fact, only 3d$_{5/2}$ and 3d$_{3/2}$ core levels at the binding energies of 573.1 and 583.5 eV, corresponding to an oxidation state Te(0), are observed in Fig.~\ref{S3} for both the pristine PtTe$_2$ (red curve) and the same surface exposed to a dose of 10$^6$ L (1 L=$1.33 \times 10^{-6}$ mbar·s) of O$_2$ at room temperature (dashed blue curve).


The Pt 4f doublet reported in Fig.~\ref{S3} is observed at binding energies of 72.5 (4f$_{7/2}$) and 75.9 (4f$_{5/2}$) eV, respectively. Similar values of the binding energy for Pt-4f core levels have been reported for PtSe$_2$ \cite{doi:10.1021/acsnano.6b04898,ADMA201602889}. 

\begin{figure}[h]
\includegraphics[width=.8\linewidth]{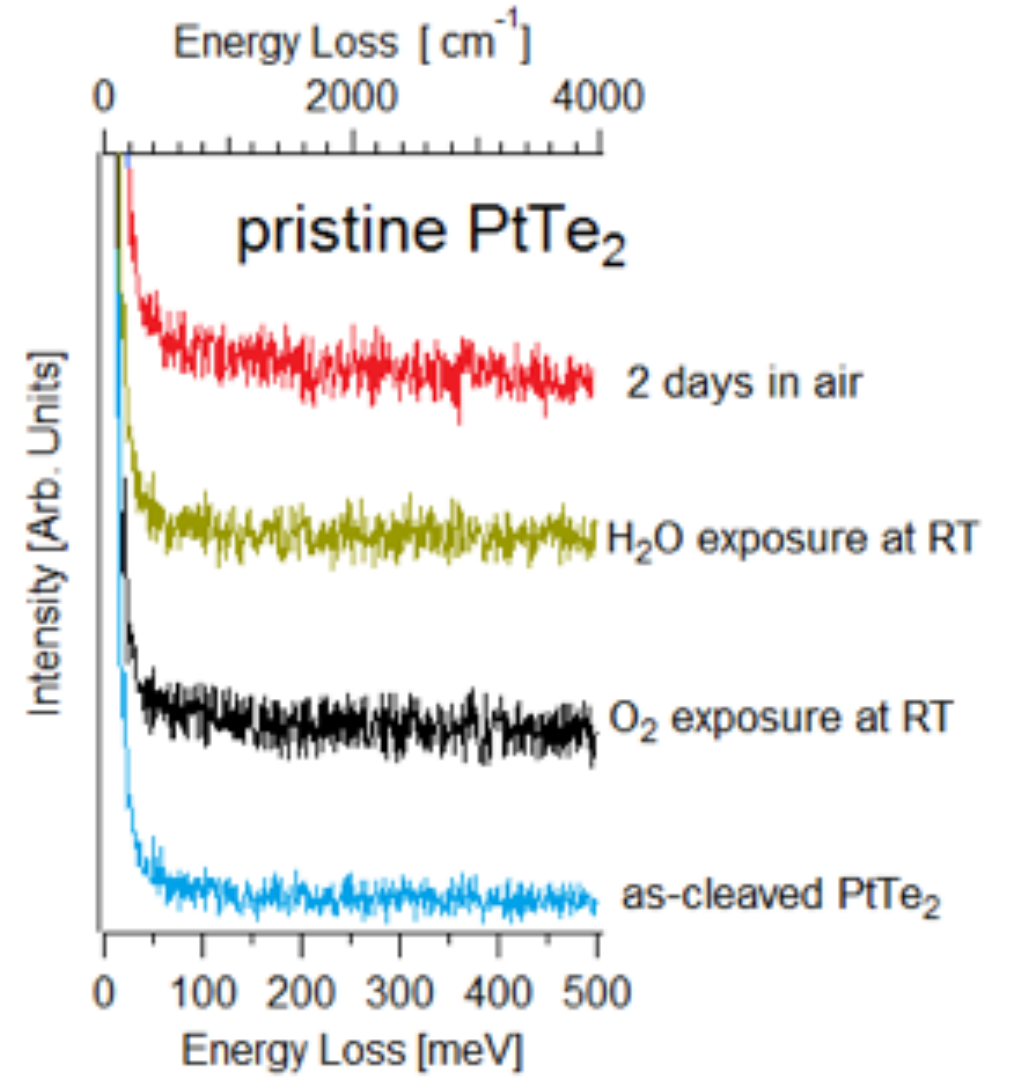}
\caption{Vibrational spectra for as-cleaved PtTe$_2$ and for the same surface exposed to $10^3$ L of O$_2$ and H$_2$O at room temperature. Successively, the surface has been exposed to air. No vibrational peak is observed.}
\label{S4}
\end{figure}

To further assess the surface chemical reactivity in PtTe$_2$-based systems, we have also carried out vibrational experiments by means of high-resolution electron energy loss spectroscopy (HREELS). Explicitly, we have exposed pristine PtTe$_2$ to water and oxygen. The corresponding vibrational spectra are reported in Fig.~\ref{S4}. For the case of pristine (undefected) PtTe$_2$, the vibrational spectrum remains featureless even after exposure of water and oxygen at room temperature (Fig.~\ref{S4}). Likewise, no vibrational peaks are revealed in the air-exposed undefected PtTe$_2$ surface (red curve in Fig.~\ref{S4}). Combined with the XPS results reported in Fig.~\ref{S3}, these data lead to the conclusion that undefected PtTe$_2$ does not react at room temperature with ambient gases. This finding has a particular interest in view of applications in optoelectronics based on PtTe$_2$. 

\bibliography{PtTe2_HEP.bib}
\end{document}